\documentclass[twocolumn, superscriptaddress, prb, footinbib]{revtex4-1}

\usepackage{graphicx}
\usepackage{amsmath}
\newcommand{\mm}{\ensuremath{\mathcal{M}}}

\begin{document}

\title{High-throughput screening of small-molecule adsorption in MOF}

\author{Pieremanuele Canepa} \affiliation{Department of Physics, Wake
Forest University, Winston-Salem, NC 27109, USA.}

\author{Calvin A. Arter} \affiliation{Department of Physics, Wake
Forest University, Winston-Salem, NC 27109, USA.}

\author{Eliot M. Conwill} \affiliation{Department of Physics, Wake
Forest University, Winston-Salem, NC 27109, USA.}

\author{Daniel H. Johnson} \affiliation{Department of Physics, Wake
Forest University, Winston-Salem, NC 27109, USA.}

\author{Brian A. Shoemaker} \affiliation{Department of Physics, Wake
Forest University, Winston-Salem, NC 27109, USA.}

\author{Karim Z. Soliman} \affiliation{Department of Physics, Wake
Forest University, Winston-Salem, NC 27109, USA.}

\author{T. Thonhauser} \email{thonhauser@wfu.edu}
\affiliation{Department of Physics, Wake Forest University,
Winston-Salem, NC 27109, USA.}

\begin{abstract}
Using high-throughput screening coupled with state-of-the-art van der
Waals density functional theory, we investigate the adsorption
properties of four important molecules, H$_2$, CO$_2$, CH$_4$, and
H$_2$O in MOF-74-\mm\ with $\mm = \text{Be}$, Mg, Al, Ca, Sc, Ti, V, Cr,
Mn, Fe, Co, Ni, Cu, Zn, Sr, Zr, Nb, Ru, Rh, Pd, La, W, Os, Ir, and Pt.
We show that high-throughput techniques can aid in speeding up the
development and refinement of effective materials for hydrogen storage,
carbon capture, and gas separation. The exploration of the
configurational adsorption space allows us to extract crucial
information concerning, for example, the competition of water with
CO$_2$ for the adsorption ``pockets.'' We find that only a few noble
metals---Rh, Pd, Os, Ir, and Pt---favor the adsorption of CO$_2$ and
hence are potential candidates for effective carbon-capture materials.
Our findings further reveal significant differences in the binding
characteristics of H$_2$, CO$_2$, CH$_4$, and H$_2$O within the MOF
structure, indicating that molecular blends can be successfully
separated by these nano-porous materials. 
\end{abstract}

\date{\today}
\maketitle

\section{Introduction}
\label{sec:intro}

The modular building-block nature of metal organic framework (MOF)
materials makes these hybrid systems very intriguing for a variety of
technologically important applications, ranging from gas storage
\cite{Valenzano10, Valenzano11, Queen11, Yao12, Nijem12, Canepa13} and
gas sequestration\cite{Millward05, Britt08, Britt09, Valenzano10,
Valenzano11, Queen11, Yao12, Nijem12, Canepa13} to more exotic
applications.\cite{Canepa13a, Uemura09, Vitorino09, Allendorf09,
White09, Tan11, Bordiga04, Kurmoo09, Horcajada09, Stroppa11,
TanK12,Li12, Nijem12a, Qiu09, Stroppa13, Serre07, Soo07, Allendorf08,
Kreno12, Xie10} The extraordinary diversity demonstrated by MOFs derive
primarily from their vast range of organic linkers combined with the
wide chemistry of metal atoms (or clusters), which alter their responses
to many external physical and chemical stimuli and influence their
flexibility, affinity towards adsorbing molecules, and intrinsic
reactivity. In this regard, much progress has been made improving the
adsorption properties of MOFs. For example, MOFs with unsaturated metal
centers such as MOF-74-\mm\ with $\mm =\text{Mg}$, Mn, Fe, Co, Ni, Cu,
and Zn show improved adsorption densities for H$_2$ and CH$_4$ and
faster adsorption at small partial CO$_2$ pressures, the latter of which
is highly desirable for CO$_2$ capturing applications.\cite{Rosweel06,
Liu08, Caskey08, Nijem10}

Although considerable experimental effort has gone into the synthesis,
characterization, and study of adsorption properties of target molecules
in MOFs, such work typically requires a significant amount of time,
slowing down scientific progress. Thus, the help of computational
material science becomes crucial, accelerating and guiding the
refinement of existing materials as well as the prediction of new MOFs.
A very promising approach is the so-called \emph{high-throughput
screening} (HTS), which---in a much shorter time compared to
experiment---screens many possible materials; it is well established in
the fields of pharmacology and biology and just recently was introduced
into the materials science community. Excellent examples of HTS are the
materials project\cite{MaterialsProject,Jain2011} and the material
genome initiative.\cite{genome}

In this work we demonstrate the importance of HTS to accelerate the
discovery of MOFs with better adsorption properties for gas-storage and
gas-separation applications.  We focus on one particular MOF, i.e.\
MOF-74, because of its unprecedented adsorption characteristics and
specificity towards CO$_2$, which makes it very important for the
process of separating CO$_2$ from CH$_4$ in low-quality gas such as
biogas.  We start with MOF-74-Zn and use HTS to study its large
\emph{configurational adsorption space} and \emph{element space}.
Specifically, we study  the adsorption properties of four important
molecules, i.e.\ H$_2$, CH$_4$, CO$_2$, and H$_2$O in combination with
25 different metals, which are: Be, Mg, Al, Ca, Sc, Ti, V, Cr, Mn, Fe,
Co, Ni, Cu, Zn, Sr, Zr, Nb, Ru, Rh, Pd, La, W, Os, Ir, and Pt.
Interestingly, from this list only eight iso-structural MOF-74-\mm\ with
$\mm = \text{Ti}$, Mg, Mn, Ni, Co, Fe, Zn, and Cu have been synthesized
since 2005, attesting to the long experimental time-scale. The
pioneering contribution of Park \emph{et al.}\cite{Park12} in the study
of CO$_2$ adsorption in MOF-74-\mm\ (with $\mm =\text{Mg}$, Ca, and the
first row of transition metals) constitutes a sub-set of our study and
serves as a benchmark.  But, while that study is limited to only CO$_2$
adsorption, we go beyond that by considerably extending the list of
possible metals and also studying adsorption of H$_2$, CH$_4$, and
H$_2$O. The effect of water is particularly important in that it is
always present in the form of humidity in flue gases and might
pre-adsorb at the unsaturated metal sites of MOF, hindering the
adsorption and transport properties of other target
molecules.\cite{Canepa13}

\section{Computational Details}
\label{sec:comp}

To explore the binding configurational space in terms of metal species
and adsorbing molecules we use density functional theory with the van
der Waals density functional vdW-DF,\cite{Dion07, Thonhauser07,
Langreth09} as implemented in VASP.\cite{Kresse93, Kresse94, Kresse96,
Kresse96a, Klimes10, Klimes11} We have already successfully applied
vdW-DF to investigate the adsorption of small molecules in MOFs and
nano-structures in numerous other studies.\cite{Yao12, Nijem12, TanK12,
Li12, Nijem12a, Canepa13, Lopez13} In particular, vdW-DF is crucial for
correctly describing the binding of water.\cite{Kolb11}

Due to the large unit cell of MOF-74 with 54 atoms, the total energy was
sampled at the $\Gamma$-point only.  Projector augmented-wave
theory,\cite{Bloch94,Kresse99} combined with a well-converged plane-wave
cutoff of 480 eV were used to describe the wave functions. The
convergence threshold for the total energy was set to 1$\times$10$^{-5}$
eV, ensuring an accurate calculation of the adsorption energies. The
internal geometry and unit cell of MOF-74-\mm\ were fully relaxed for
all \mm\ using vdW-DF \cite{Sabatini12} (empty and fully loaded with
H$_2$, CO$_2$, H$_2$O, and CH$_4$) until the force and stress criteria
of 1$\times$10$^{-3}$~eV~\AA$^{-1}$ and 1$\times$10$^{-3}$~eV~\AA$^{-3}$
were satisfied.  For the study of the electronic structure of these MOF
materials we carried out the Bader analysis using the fast
implementation proposed by Henkelman \emph{et al.}\cite{Henkelman2006}
Graphical manipulations were carried out using
\emph{J-ICE}.\cite{Canepa11a}

Our calculations start from the experimental hexagonal structure of
MOF-74-Zn with space group $R\overline{3}$ and $a = 25.932 $~\AA\ and $c
=6.836$~\AA\ and 54 atoms per unit cell.\cite{Rosi05} We then swap out
the Zn atoms in sequence with Be, Mg, Al, Ca, Sc, Ti, V, Cr, Mn, Fe, Co,
Ni, Cu, Sr, Zr, Nb, Ru, Rh, Pd, La, W, Os, Ir, and Pt, respectively.
Originally, we also considered Y, Mo, Ag, Ce, and Au, however, their
respective MOF structures are relatively unstable preventing their
convergence. Since some of these metals present an open-shell electronic
structure, we adopted a collinear spin-corrected treatment, with an
appropriate approximation for the vdW-DF part.\cite{Kolb13} We impose an
anti-ferromagnetic alignment of the spins on the six metal ions in the
unit cell, according to previous experimental\cite{Dietzel08} and
theoretical\cite{Canepa13a} observations. Six H$_2$,\cite{Liu08}
CO$_2$,\cite{Dietzel08} H$_2$O,\cite{Dietzel06} and CH$_4$\cite{Wu09}
molecules per unit cell are then adsorbed at the uncoordinated metal
sites \mm\ in the MOF nano-pores, reproducing scenarios of channel
saturation of previous X-ray and neutron-diffraction
experiments.\cite{Rosi05}

\section{Results}
\label{sec:results}

\subsection{Properties of the Empty MOF}

We begin our discussion by commenting on the structural characteristics
of the empty MOF-74.  Table~\ref{table:dataMOF} shows the structural
parameters and other relevant quantities of MOF-74 after complete
structural relaxation at 0 K.  There is not a simple explanation for the
dependence of the lattice parameters on the metal species. But, it is
interesting to see that Os results in the smallest unit cell, while Ca
results in the largest. The significant difference between them can be
associated to the ionic radii of Ca$^{2+}\approx 1.00$ \AA\ and
Os$^{2+}\approx 0.49$ \AA\ ions when in a penta-coordinated oxygen
environment (as found in MOF-74-\mm).\cite{Shannon76} Note that the
other metal species do not necessarily follow such a simple trend, and
hence we cannot extrapolate a general dependence of ionic radius
\emph{vs.} volume.

We also simulated the powder X-ray spectra of a few selected MOFs
obtained throughout the HTS procedure, resulting in important
fingerprints for their future synthesis (see Supplementary Materials).

\begin{table}[t]
\caption{\label{table:dataMOF} Computed lattice constants $a$ and $c$
(in \AA) and volume $V$ (in \AA$^3$). Atomic numbers \emph{Z}, and Bader
charges $Q_\mm$ (in units of $e$) at the metal sites \mm\ are also
reported.} 
\begin{tabular*}{\columnwidth}{@{\extracolsep{\fill}}lccccc@{}}\hline\hline
\mm &  \emph{Z}  &\emph{a} & \emph{c} & \emph{V} &  $Q _\mm$ \\\hline
Be &  4  & 25.655 & 6.663 & 3797.877 & 1.6 \\
Mg &  12 & 26.084 & 6.863 & 4043.947 & 1.5 \\
Al &  13 & 25.402 & 6.565 & 3668.630 & 2.6 \\
Ca &  20 & 25.454 & 7.591 & 4259.190 & 1.5 \\
Sc &  21 & 23.675 & 7.334 & 3559.960 & 1.9 \\
Ti &  22 & 23.669 & 7.210 & 3498.429 & 1.8 \\
V  &  23 & 25.254 & 7.000 & 3868.982 & 1.6 \\
Cr &  24 & 26.171 & 6.525 & 3870.148 & 1.5 \\
Mn &  25 & 26.242 & 7.082 & 4223.524 & 1.4 \\
Fe &  26 & 26.010 & 6.711 & 3931.742 & 1.3 \\
Co &  27 & 26.078 & 6.872 & 4047.173 & 1.3 \\
Ni &  28 & 25.688 & 6.188 & 3536.291 & 1.1 \\
Cu &  29 & 26.271 & 6.138 & 3668.332 & 0.8 \\
Zn &  30 & 26.142 & 6.875 & 4068.779 & 1.2 \\
Sr &  38 & 26.683 & 6.710 & 4137.427 & 1.6 \\
Zr &  40 & 23.455 & 7.530 & 3587.630 & 2.0 \\
Nb &  41 & 27.031 & 6.414 & 4058.779 & 1.4 \\
Ru &  44 & 27.061 & 6.119 & 3880.592 & 1.3 \\
Rh &  45 & 25.833 & 6.804 & 3932.355 & 1.3 \\
Pd &  46 & 26.570 & 6.432 & 3932.482 & 1.1 \\
La &  57 & 26.672 & 6.431 & 3962.091 & 2.2 \\
W  &  74 & 26.960 & 6.177 & 3888.314 & 1.6 \\
Os &  76 & 26.480 & 4.977 & 3022.272 & 1.8 \\
Ir &  77 & 26.020 & 6.796 & 3984.552 & 1.2 \\
Pt &  78 & 26.560 & 6.511 & 3977.779 & 1.2 \\
\hline\hline
\end{tabular*}
\end{table}

\begin{figure}[t]
\includegraphics[width=\columnwidth]{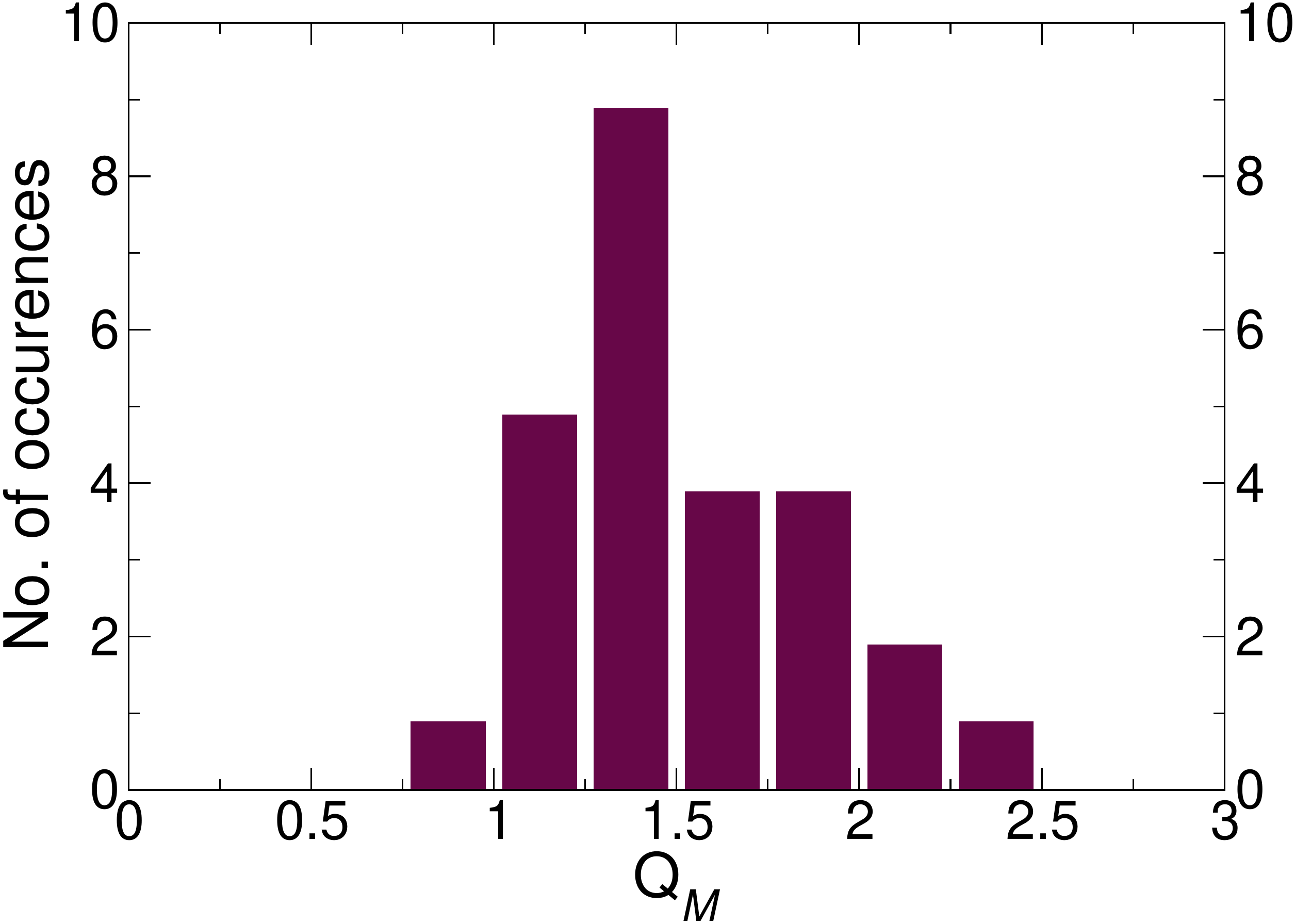}
\caption{\label{fig:chargehist} Number of occurrences of particular
$Q_\mm$ in Table~\ref{table:dataMOF}.}
\end{figure}

It is also interesting to study the Bader charges of the metal ions in
the MOF. Figure~\ref{fig:chargehist} analyzes the 25 situations in
Table~\ref{table:dataMOF} and plots the number of occurrences of Bader
charges $Q_\mm$.  From this figure we see that most of the metal species
in MOF-74 display charges ranging from 1.0 to 2.0~$e$, consolidating the
picture of divalent metal ions. Exceptions are Al, which  carries almost
a 3+ charge as expected, and Cu, which remains as Cu(I).  Our finding
also suggest that Rh, Pd, Os, Ir, and Pt remain weakly charged
preserving their noble metal characteristics.  The local oxygen
environment experienced by the metal species of MOF-74 resembles a
``surface termination'' of the corresponding binary oxides, thus
explaining the charge nature of these ions.  Note that the charge
characteristics of such metal ions reflects their reactivity towards the
adsorbates.

\subsection{Adsorption Characteristics}

Our discussion now moves to the analysis of the adsorption energies,
which determine if molecular adsorption is favorable. We define the
adsorption energy as in Eq.~(\ref{eq:be})
\begin{equation}
\label{eq:be}
\Delta E = E_{\rm MOF+M} - E_{\rm MOF} - E_{\rm M}{\rm(g)}\;,
\end{equation}
where $E_{\rm MOF+M}$, $E_{\rm MOF}$, and $E_{ \rm M}$\rm{(g)} are the
total energy of MOF with molecules adsorbed in its nano-pore, the energy
of the empty MOF, and the energy of the molecule in its gas phase
geometry, respectively. Note that throughout the manuscript we will also
refer to the adsorption energies as binding energies. Two interesting
deformation contributions $\delta E _{\rm M}$ and $\delta E_{\rm MOF}$,
which are clearly connected to the adsorption process, are defined in
Eqs.~(\ref{eq:sdelta}) and (\ref{eq:sdeltamof}):
\begin{eqnarray}
\delta E _{\rm M}  &=& E_{\rm M,\,in\,MOF+M} - E_{\rm M}{\rm(g)}\;,
\label{eq:sdelta}\\ 
\delta E _{\rm MOF} &=& E_{\rm MOF,\,in\,MOF+M} - E_{\rm MOF}\;,
\label{eq:sdeltamof}
\end{eqnarray}
where $E_{\rm M,\,in\,MOF+M}$ and $E_{\rm MOF,\,in\,MOF+M}$ are the
energies of the molecule and the MOF in their adsorption geometries.
$\delta E _{\rm M}$ and $\delta E_{\rm MOF}$ express the cost in energy
that both adsorbate and MOF have to pay during the adsorption process.
The $\delta E _{\rm M}$ term also contains the lateral interactions
between adjacent molecules, which in turn depend on the electronic
characteristic of each molecule and their mutual positions. Note that
both $\delta E_{\rm M}$ and $\delta E_{\rm MOF}$ contributions are
related to the re-arrangement of the molecular and MOF geometries in
order to maximize the binding interaction.  $\delta E_{\rm M}$ and
$\delta E_{\rm MOF}$ are obtained by partitioning the adsorption energy,
and thus they are naturally enclosed in the definition of $\Delta E$ in
Eq.~(\ref{eq:be}). Finally, it is possible to define an adsorption
quantity $\Delta E^{\rm C}$ free of any deformation contributions as
\begin{equation}
\label{eq:bec}
\Delta E^{\rm C} = \Delta E - \delta E _{\rm M} - \delta E _{\rm MOF}\;.
\end{equation}
Table~\ref{table:data} is the main result of this paper and collects the
calculated values for the quantities defined in Eqs.~(\ref{eq:be}) --
(\ref{eq:bec}).

\begin{table*}
\renewcommand{\tabcolsep}{1mm}
\scriptsize
\caption{\label{table:data} Computed adsorption energies $\Delta E$ and
derived quantities $\Delta E^{\rm C}$, $\delta E_{\rm MOF}$, and $\delta
E _{\rm M}$ (in kJ~mol$^{-1}$), for MOF-74-\mm\ with adsorbed H$_2$,
CO$_2$, CH$_4$, and H$_2$O molecules. $\Delta E$ and other contributions
are reported per adsorbed molecules. } 
\begin{tabular*}{\textwidth}{@{}l|@{\extracolsep{\fill}}cccc|cccc|cccc|cccr@{}}
\hline\hline  
 & \multicolumn{4}{c|}{H$_2$} &  \multicolumn{4}{c|}{CO$_2$}  &
\multicolumn{4}{c|}{CH$_4$} &  \multicolumn{4}{c}{H$_2$O} \\
\mm &    $\Delta E$ &  $\Delta E^{\rm C}$ & $\delta E_{\rm MOF}$ &$\delta E _{\rm M}$ 
   & $\Delta E$ &  $\Delta E^{\rm C}$ & $\delta E_{\rm MOF}$ &$\delta E _{\rm M}$ 
   & $\Delta E$ &  $\Delta E^{\rm C}$ & $\delta E_{\rm MOF}$ &$\delta E _{\rm M}$ 
   & $\Delta E$ &  $\Delta E^{\rm C}$ & $\delta E_{\rm MOF}$ &$\delta E _{\rm M}$ \\
\hline
Be & --16.5 & --15.1 &  0.4  & --0.8 & --60.2 & --47.1 & 4.2 & --18.1 & --40.2 & --35.3 & 0.9  & --5.8 &  --41.4  & --39.7 &  2.3  & --4.0  \\
Mg & --15.8 & --15.5 &  0.3  & --0.7 & --48.2 & --46.4 & 1.2 & --3.0  & --37.0 & --36.0 & 1.3  & --4.6 &  --73.2  & --72.5 &  4.2  & --5.0  \\
Al & --19.8 & --18.6 & --0.9 & --0.3 & --118.4& --452.6&126.6& 207.7  & --38.2 & --34.5 & 1.0  & --4.6 &  --135.7 & --158.2&  23.6 & --1.1  \\
Ca & --18.7 & --19.5 & 1.2   & --0.4 & --57.0 & --49.0 & 1.7 & --10.0 & --40.1 & --35.7 & 0.8  & --5.2 &   --87.1 & --93.9 &  9.1  & --2.2  \\
Sc & --19.6 & --19.6 &  0.6  & --0.6 & --53.0 & --45.9 & 1.5 & --8.5  & --45.0 & --39.6 & 1.3  & --7.0 &  --113.1 & --128.8&  16.5 & --0.9  \\
Ti &   1.4  & --18.6 & 12.0  & 8.0   & --49.4 & --41.8 & 5.0 & --12.6 & --39.9 & --36.1 & 5.8  & --9.6 &  --50.7  & -20.9  &  6.0  & --35.7 \\
V  & --20.0 & --19.0 & --0.6 & --0.4 & --52.7 & --59.5 & 7.5 & --1.0  & --43.3 & --42.5 & 2.4  & --3.2 &  --110.9 & --116.1&  6.1  & --0.9  \\
Cr & --19.8 & --19.6 &  0.2  & -0.5  & --52.9 & --49.5 & 1.0 & --4.4  & --37.8 & --33.1 & 1.2  & --5.9 &  --51.1  & --61.1 &  12.4 & --2.4  \\
Mn & --19.0 & --19.1 &  0.5  & -0.3  & --53.7 & --52.8 & 2.6 & --3.6  & --43.2 & -38.9  & 0.5  & --4.9 &  --73.1  & --81.8 &  11.3 & --2.7  \\
Fe & --19.8 & --19.1 & --0.2 & --0.4 & --51.2 & --47.2 & 1.4 & --5.4  & --39.8 & --35.4 & 1.5  & --5.9 &  --129.7 & --163.8&  30.1 & --4.0  \\
Co & --19.8 & --19.5 &  0.5  & --0.4 & --40.8 & --36.1 & 1.5 & --5.4  & --37.4 & --38.1 & 0.8  & --5.1 &  --71.7  & --80.8 &  12.1 & --3.0  \\
Ni & --19.1 & --18.0 &  0.3  & --1.5 & --41.4 & --35.7 & 0.6 & --6.1  & --36.0 & --33.1 & 0.6  & --6.3 &  --60.6  & --61.2 &  3.9  & --3.4  \\
Cu & ---$^a$& ---    & ---   & ---   & --42.9 &  --43.1& 0.8 & --0.6  & --39.7 & --35.2 & 0.6  & --5.1 &  --90.3  & --80.9 & --18.1&   8.8  \\
Zn & --20.5 & --19.4 &  0.4  & --2.0 & --52.4 & --46.4 & 1.0 & --3.1  & --44.6 & --40.1 & 1.7  & --5.8 &  --75.5  & --75.2 &  3.4  & --3.8  \\
Sr & --18.6 & --19.4 &  1.2  & --0.4 & --49.7 & --53.2 & 5.2 & --1.7  & --43.9 & --38.6 &-0.2  & --5.1 &  --153.6 & --185.7& 31.2  &   0.8  \\
Zr & --17.8 & --17.1 &  0.2  & --0.9 & --52.0 & --41.8 & 1.7 & --11.9 & --43.8 & --37.6 & 2.1  & --8.4 &  --90.3  & --116.7&  29.7 & --3.3  \\
Nb & --20.7 & -20.7  & --0.2 & 0.2   & --89.1 & --94.1 & 5.5 & --0.6  & --44.5 & --48.4 & 4.9  & --1.0 &  --124.5 & --126.3&  3.6  & --1.8  \\
Ru & --20.5 & --19.5 & --0.3 & --0.7 & --49.5 & --48.8 & 1.8 & --2.5  & --38.5 & --36.3 & 1.7  & --3.8 &  --77.5  & --61.7 &  15.0 & --0.1  \\
Rh & --20.8 & --19.9 & --0.5 & --0.4 & --52.5 & --46.7 & 1.3 & --7.1  & --36.1 & --34.6 & 1.0  & --6.2 &  --50.5  & --48.1 &  0.6  & --2.9  \\
Pd & --19.5 & --19.0 & 0.1   & --0.7 & --51.3 & --49.3 & 0.2 & --2.3  & --37.4 & --32.3 & 0.6  & --5.7 &  --46.1  & --44.2 &  0.7  & --2.6  \\
La & --20.2 & --18.7 &--0.9  & --0.7 & --90.0 & --102.8& 6.2 &  6.6   & --40.9 & --36.9 & 0.8  & --4.9 &  --105.2 & --122.6&  17.1 &   0.3  \\
W  & --21.9 & --20.6 &--0.9  & --0.4 & --52.0 & --47.2 & 0.5 & --5.2  & --40.8 & --48.2 & 1.7  & 5.6   &  --133.2 & --141.3&  7.9  &   0.2  \\
Os & --19.1 & --9.4  &  0.3  & --10.0& --58.8 & --43.9 & 1.1 & --16.2 & ---$^a$& ---    &  --- & ---   &  --50.5  & --53.4 &  4.7  & --1.7  \\
Ir & --20.4 & --19.5 & --0.4 & --0.4 & --55.1 & --45.9 & 8.8 & --18.0 & --36.8 & --33.1 & 1.2  & --6.1 &  --49.0  & --46.7 &  1.2  & --2.8  \\
Pt & --19.3 & --18.4 & --0.4 & --0.6 & --52.2 & --49.9 & 0.2 & --2.5  & --36.1 & --31.2 & 1.0  & --5.9 &  --45.1  & --43.2 &  1.0  & --2.4  \\
\hline
\hline
\end{tabular*}
$^a$Simulation considered not converged since we observe unphysical
molecular dissociations due to huge structural strains.
\end{table*}

\begin{figure}[t]
\includegraphics[width=\columnwidth]{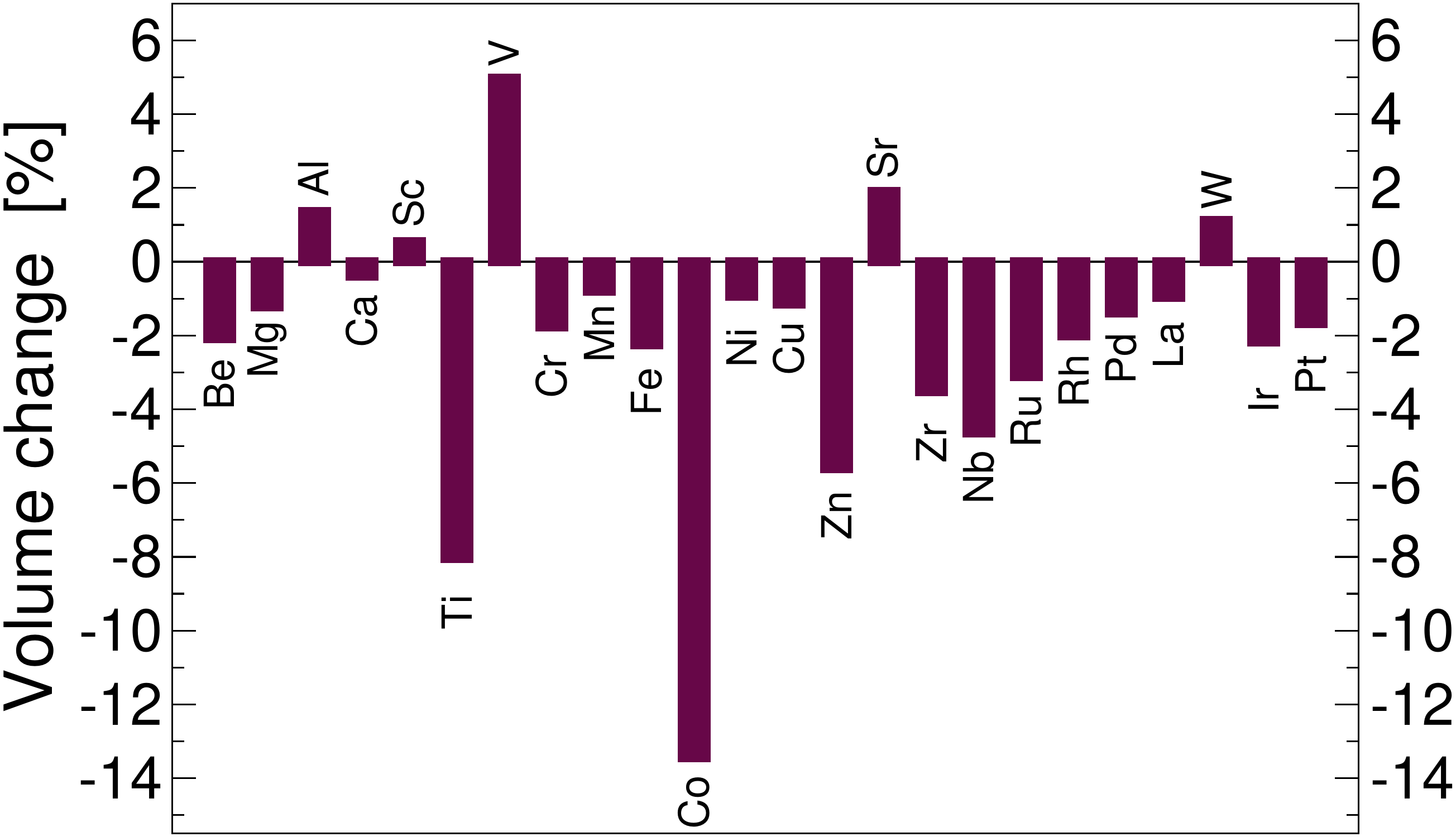}
\caption{\label{fig:expansion} Relative volume change (in \%) of
MOF-74-\mm\ after six CH$_4$ molecules have been introduced into its
cavity.}
\end{figure}

We begin by making some observations on the evolution of the structures
of these MOF-74-\mm\ once the adsorbing molecules is introduced.  A
first glance tells us that the pore size and volume of the MOFs decrease
when molecules are adsorbed in their structures (see Table~S1 in the
Supplementary Materials). The extent of volume change is balanced
between the size of the adsorbing molecules and the nature of the
adsorption interactions in Table~\ref{table:data}. For example, we
observe the ``clog up'' of the MOF nano-pore when six CH$_4$ molecules
are concomitantly adsorbed, hence increasing the lateral molecule to
molecule interaction as demonstrated by the $\delta E_M$ in
Table~\ref{table:data}. The presence of large, attractive $\delta E_M$
values contributes to ``unphysically'' lowering the overall $\Delta E$.
Figure~\ref{fig:expansion} shows the perturbation introduced in the
volume of MOF-74-\mm\ after six CH$_4$ molecules are introduced in its
cavity.  The trend in volume changes is not directly correlated to
$\Delta E$, but rather to $\delta E_M$ (see Table~\ref{table:data}),
which combines the molecular deformations and the lateral interactions
experienced by the molecules during the adsorption. In the case of
CH$_4$ the intermolecular interactions are considerably stronger and
keep the MOF structure from swelling apart, which could occur otherwise
given the substantial volume of six CH$_4$ molecules in such a small
pore (diameter $\sim$13~\AA). Similar conclusions can also be drawn for
H$_2$, CO$_2$, and H$_2$O, however, the effect on the respective volume
changes is smaller.  The different changes in volume due to the diverse
nature of the MOF/molecule interaction and the molecular volume again
reflect the high structural flexibility displayed by these porous
materials. In other words, the MOF structure responds differently to
different molecules, manifesting significant and unprecedented molecular
recognition effects that could be exploited for sensor applications and
are investigated in a forthcoming publication.

The diverse molecular recognition displayed by these MOFs is also
revealed  by the binding energies $\Delta E$ of H$_2$, CO$_2$, CH$_4$,
and H$_2$O in MOF-74-\mm\ (see Table~\ref{table:data}). In general, we
find that the $\Delta E$s of these molecules are substantially different
in magnitude by tens of kJ~mol$^{-1}$ and follow a precise trend H$_2$O
$\gg$ CO$_2$ $>$ CH$_4$ $\gg$ H$_2$. The consistent gap between the
adsorption energies of different molecules indicates that such MOFs can
be efficiently used in filters for the separation of blended gases. 

According to the electrostatic complementarity principle
(donor-acceptor), the metal species of the MOF structures act as Lewis
acid laking in electrons, whereas the adsorbing counterparts (in general
oxygen atoms) behave as Lewis base providing electrons. It is well
established that in the case of H$_2$O the driving force dictating the
molecular adsorption at the metal site is its strong dipole moment,
whereas the adsorption of H$_2$, CO$_2$, and CH$_4$ relies on weak van
der Waals forces.  This is demonstrated by the $\Delta E$ of
Table~\ref{table:data}. Water clearly remains the preferred molecule for
the metal sites, with the exception of some noble metals i.e. Rh, Pd,
Os, Ir, and Pt (see below), implying that performance for gas-storage
applications can be hindered in moisturized environments. The presence
of water in these nano-porous materials is scarcely documented in the
literature,\cite{Nachtigall10, Valenzano12, Robinson12,Canepa13} but
remains a major operational problem; it is partly for this reason that
non hydro-soluble MOFs such as fluorinated MOFs are being
developed.\cite{Yang07,Yang09}

It is also interesting to look at the different adsorption energy
contributions, i.e.\ $\delta E_{\rm M}$, $\delta E_{\rm MOF}$, and
$\Delta E^{\rm C}$ in Table~\ref{table:data}. As anticipated, the
negative sign of the $\delta E_{\rm M}$ is simply due to the attractive
intermolecular interactions and their magnitudes only depend on the
molecular size and the extent of pore reconstruction, the latter being
connected to the nature of the metal ions. On the other hand, the sign
of the $\delta E_{\rm MOF}$ is always positive (with very few
exceptions) and is thus an indication that the MOF undergoes an
unfavorable reconstruction when the molecule is adsorbed.  In the
majority of the cases the $\delta E_{\rm M}$ is larger in magnitude than
$\delta E_{\rm MOF}$.

In general, the calculated $\Delta E$s of Table~\ref{table:data} are in
good agreement with previous experimental and computational
data.\cite{Nijem10, Valenzano10, Valenzano11, Valenzano12, Yao12,
Nijem12, Park12, Canepa13} Our data reproduces the $\Delta E$ order
established for CO$_2$ in MOF-74-Mg $>$ MOF-74-Ni $>$ MOF-74-Co,
consistent with previous experimental investigations.\cite{Park12} The
adsorption energies calculated for most of the molecules are slightly
overestimated by 2 -- 5 kJ~mol$^{-1}$ from the experimental data (where
available), which is typical for the vdW-DF
functional.\cite{Thonhauser06} Nevertheless, inclusion of the zero-point
energy and temperature effects \emph{via} phonon calculations (not
performed in this study) can lower the $\Delta E$s and bring them in
better agreement with experiment.\cite{Valenzano10, Valenzano11,
Valenzano12, Canepa13} $\Delta E^{\rm C}$, rather than $\Delta E$, is
more appropriate for comparison with experimental data, as it is not
affected by the spurious lateral interactions introduced by the
high-loading regimes imposed in our simulations (six molecules per
cell).

\subsection{Adsorption of H$_2$ and CH$_4$}

Somewhat surprisingly, the data in Table~\ref{table:data} demonstrates
that for most cases the adsorption energies of H$_2$ in MOF-74 only
marginally change with the metal species. It follows that the currently
synthesized MOF-74-Mn, MOF-74-Fe, MOF-74-Co, MOF-74-Ni, and MOF-74-Zn
are already ``as good as it gets'' for the purpose of hydrogen storage.
A loading of six molecules in MOF-74-\mm\ per unit cell corresponds to a
hydrogen-storage capacity of 1.6~mass\% and
4.9~g~H$_2$~L$^{-1}$.\cite{Wong-Foy06} Although not investigate here,
secondary binding sites exist in MOF-74 and the unit cell can hold 12
H$_2$ molecules under high pressure,\cite{Liu08, Lopez13} corresponding
to theoretical values of 3.2~mass\% and 9.9~g~H$_2$~L$^{-1}$ volumetric
uptake. These numbers are in the mid-range of physisorption-based
nano-porous hydrogen-storage materials.\cite{Yang10} Again, co-presence
of H$_2$O in the MOF environment is a problem for hydrogen-storage
applications, as it degrades the storage capacity even more due to the
large binding energies of water compared to H$_2$. Note that the sign of
$\Delta E$ for H$_2$ in MOF-74-Ti suggests that this adsorption is
thermodynamically prohibited.

\begin{figure}[t]
\includegraphics[width=\columnwidth]{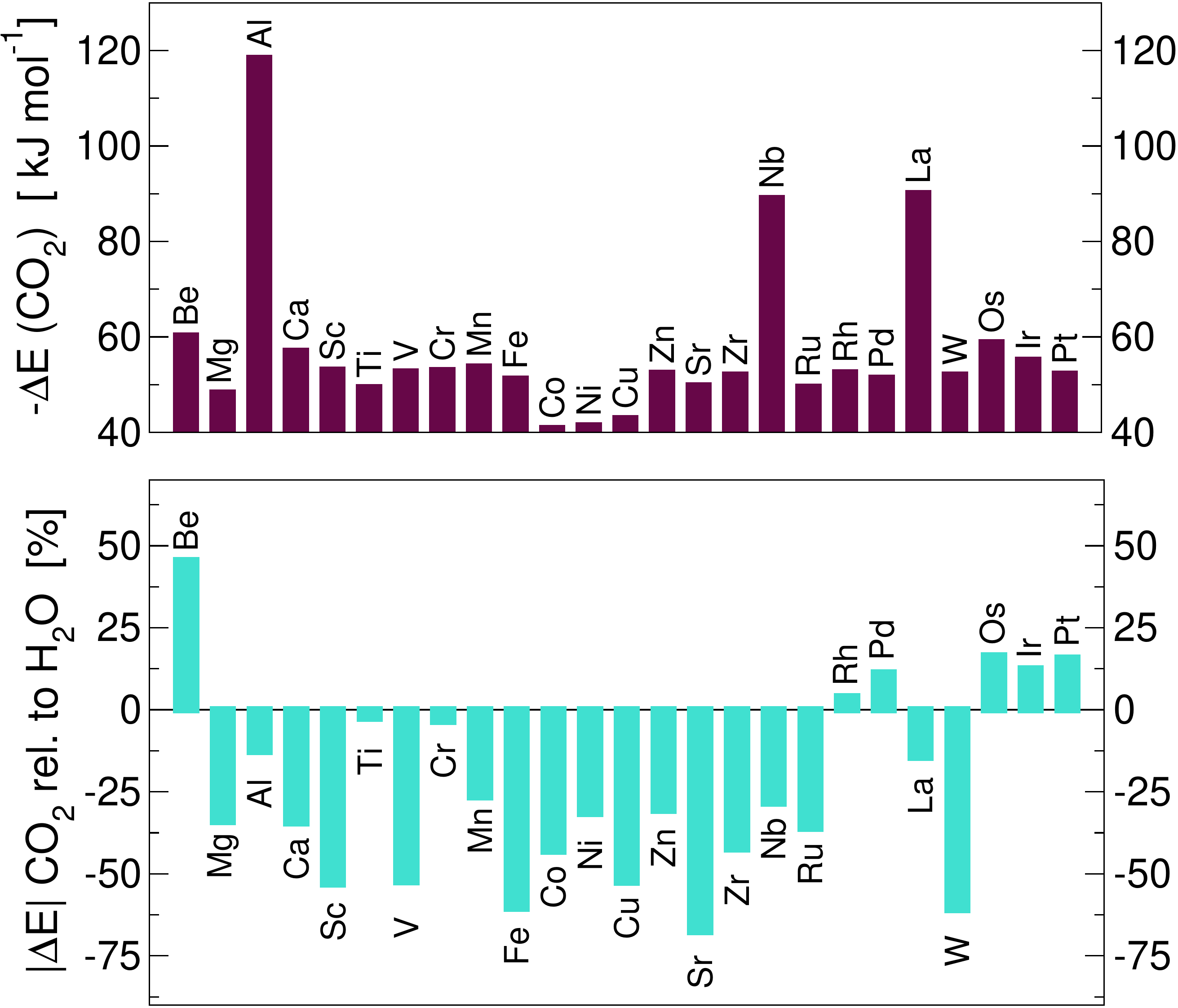}
\caption{\label{fig:co2diff} (top) $\Delta E$ for CO$_2$ adsorption (in
kJ~mol$^{-1}$) in MOF-74-\mm. (bottom) Magnitude of the adsorption
energy of CO$_2$ relative to H$_2$O. A positive value in this plot means
that CO$_2$ binds stronger than H$_2$O.}
\end{figure}

Several computational studies have investigated the interaction of
CH$_4$ with MOF-74, often using plain LDA functionals which incorrectly
describe the dominating van der Waals interactions relevant for the
adsorption.\cite{Wu09} Note that here we employ an exchange-correlation
functional that is not parametrized and hence capable of capturing the
diverse nature of the molecular interactions with all metals.  On the
other hand, the empirical method DFT+D\cite{Grimme06} works well only
for a few metals where the empirical parameters (the $C_6$ coefficients)
are extracted from \emph{ab initio} data; this is not appropriate for
metals whose $C_6$ coefficients are extrapolated from lighter elements
along the group.\cite{Civalleri08,Canepa12b} This would explain why Park
\emph{et al.}\cite{Park12} obtained very similar $\Delta E$ for the
metal ions of the first transition metal row.  Furthermore, the $C_6$
obtained by Grimme \emph{et al.}\cite{Grimme06} were derived for atomic
species and not ions, hence not reflecting the nature of the metal ions
in MOF-74 as demonstrated by the Bader charges in
Table~\ref{table:dataMOF} and Fig.~\ref{fig:chargehist}.

\subsection{Competition of H$_2$O and CO$_2$ Adsorption}

Our analysis now moves to the comparison of the computed binding
energies for CO$_2$ and H$_2$O. The top panel of Fig.~\ref{fig:co2diff}
shows that in most cases the $\Delta E$s for CO$_2$ in MOF-74-\mm\
oscillate between 40 and 60~kJ~mol$^{-1}$.  However, when adsorbing
CO$_2$ in MOF-74-Al, MOF-74-La, and MOF-74-Nb, we observe a complete
chemi-adsorption of CO$_2$ at the metal site, sharing electrons with the
MOF structure and therefore causing a steep increase of the adsorption
energies. From a practical point of view, note that strong
chemi-adsorption prevents the re-use of MOFs as the molecule is now
fully integrated in the MOF structure and its chemical identity is
unrecoverable.  In the Supplementary Materials we show the irreversible
structural and molecular changes occurring in MOF-74-Al and MOF-74-La
loaded with CO$_2$.

Very interesting are also MOF-74-Be, -Ca, and -Cr, in which we observe a
complete desorption of CO$_2$ molecules from the metal sites despite the
considerable adsorption energies reported in
Fig.~\ref{fig:co2diff}---MOF-74-Be is thus a special case and we do not
list it together with the other nobel metals that have strong affinity
towards CO$_2$. To quantify this effect, we report the molecule/metal
distances in Table S1 of the Supplementary Materials. The small affinity
of CO$_2$ for the metal species (i.e.\ Be, Ca, and Cr) causes the
molecules to reorganize close to each other, establishing strong
attractive intermolecular interactions, decreasing the overall $\delta E
_{\rm M}$ (see the Supplementary Materials).  As mentioned before, large
values of $\delta E_{\rm M}$ spuriously affect the final magnitude of
$\Delta E$, whereas the $\Delta E^{\rm C}$ are more appropriate as
reference for further comparisons.  Similar (but smaller) is the
behavior of CO$_2$ with some metal ions such as Ti, Zr, and W.

The bottom panel of Fig.~\ref{fig:co2diff} shows the binding energy of
CO$_2$ relative to that of H$_2$O. In the majority of the metals
investigated, water binds stronger than CO$_2$, reaching occasionally
more than 100~kJ~mol$^{-1}$, see for example MOF-74-Al, \mbox{-Sc}, -V,
-Fe, -Sr, -Nb, -La, and -W. A recent attempt of increasing the CO$_2$
affinity (compared to water) was proposed by Planas \emph{et
al.,}\cite{Planas13} functionalizing the metals species of MOF-74-Zn
with amines with the effect of increasing the adsorption energy of
CO$_2$. However, the reactivity of water was not tested. A closer look
at the adsorption geometries for these models explains the reasons of
such high $\Delta E$ values---we observe the incipient formation of
strong hydrogen bonds with oxygen atoms that coordinate the metal ions,
thus explaining the large structural deformation subsequent to
adsorption (see $\delta E_{\rm MOF}$ in Table~\ref{table:dataMOF}).  For
these latter cases we have plotted the density of states (DOS) of both
MOF and water molecules before and after the adsorption, demonstrating
the nature of the chemi-adsorption (see the Supplementary Materials).
Although from the DOS we notice the injection of some molecular states
in the band structure of the MOF, the geometry of the adsorption
conformation does not assist the water dissociation, in contrast to what
is largely observed for surfaces of the respective metal-oxides.

\begin{figure}[t]
\includegraphics[width=\columnwidth]{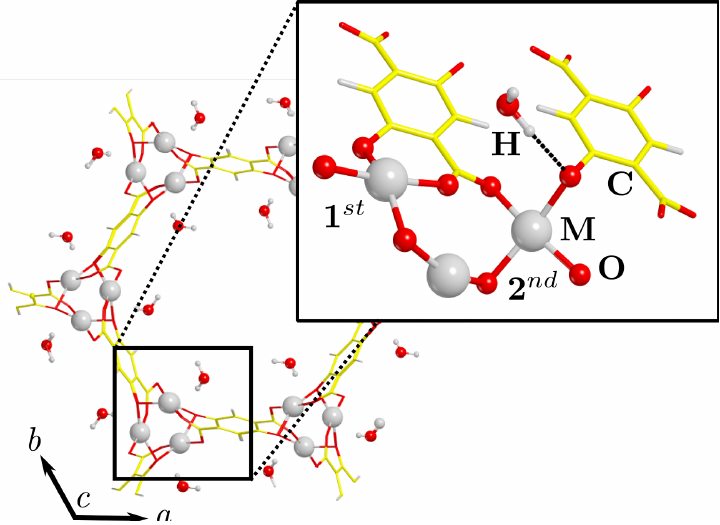}
\caption{\label{fig:site} MOF-74-\mm, with \mm\ one of the noble metals
Rh, Pd, Os, Ir, and Pt. The inset magnifies the binding site, where
1$^\text{st}$ represents the most exposed metal site, and 2$^\text{nd}$
refers to the less exposed metal site. The dashed line indicates the
hydrogen bond.}
\end{figure}

As mentioned, the noble metals Rh, Pd, Os, Ir, and Pt are special in
that they invert the trend of the adsorption energies for CO$_2$ and
H$_2$O.  The initial adsorption conformation for these cases consists of
water molecules in contact with the metal species (i.e.\ Rh, Pd, Os, Ir,
and Pt). This situation is immediately disrupted due to the redox nature
of those metals. For these cases, the noble metals act as donor
competing with the oxidizing oxygen of water leading to unfavorable
interactions, \emph{i.e.} repulsions. After complete structural
relaxation, the oxygen atoms of H$_2$O molecules are found far from the
noble metal.  Water molecules are entirely displaced from the original
binding pocket, assuming a new binding conformation that favors the
formation of hydrogen bonding with atoms of the linkers (see
Fig.~\ref{fig:site}).  We did not observe strong interactions between
the protons of H$_2$O molecules and the metal sites, a situation that is
extremely favorable in simpler systems such as water on platinum
surfaces and other noble metals.\cite{Ludwig03,Ranea04,Tatarkhanov09} To
this end, the analysis of the Bader charges of the metal ions explains
intuitively the nature of the molecule-metal interaction.  Once the
water molecules enter in contact with MOF-74-\mm\ (with $\mm =
\text{Rh}$, Pd, Os, Ir, and Pt) we find that the charge of the metal ion
is unaltered since water molecules do not directly perturb the ion.
Although for Rh, Pd, Os, Ir, and Pt the $\Delta E$s of CO$_2$ adsorption
are always larger than the ones for H$_2$O, we find that the CO$_2$
molecules also remain slightly separated from the metal ions (see the
corresponding distances in the Supplementary Materials) compared to
other transition metals, demonstrating that such noble metals prefer a
reducing environment.

\section{Conclusions}

In this study we sample the adsorption configurational space of small
molecules in MOF-74. In particular, we utilize high-throughput screening
to investigate the adsorption properties of H$_2$, CH$_4$, CO$_2$, and
H$_2$O in MOF-74-\mm\ with $\mm = \text{Be}$, Mg, Al, Ca, Sc, Ti, V, Cr,
Mn, Fe, Co, Ni, Cu, Zn, Sr, Zr, Nb, Ru, Rh, Pd, La, W, Os, Ir, and Pt.
We demonstrate that HTS can reveal important information about these
systems, which can aid in accelerating the engineering and improving of
existing metal organic frameworks for hydrogen storage, carbon capture,
and gas-separation.

Independently of the metal species of \mbox{MOF-74-\mm}, we find a
consistent gap between the adsorption energies of different molecules,
i.e.\ from strongest to weakest \ H$_2$O $\gg$ CO$_2$ $>$ CH$_4$ $\gg$
H$_2$, thus concluding that these materials can be efficiently used in
filters for the separation of blended gases. Furthermore, H$_2$O is
always present in the form of humidity in the operational environment of
MOFs, and we find that it can significantly decrease the adsorption and
transport properties of target molecules.  We further find that metal
species at the left of the periodic table are less effective in
capturing CO$_2$, displaying a larger affinity for H$_2$O---an
indication that these MOFs are susceptible to moisturized environments.
Our analysis suggests an improving affinity for CO$_2$ when moving
towards the right along the transition metal series.  On the other hand,
our data does not suggest a systematic trend along each group. The redox
characteristics of noble metals such as Rh, Pd, Os, Ir, and Pt in MOF-74
increase the interaction with CO$_2$, while the affinity for water is
almost suppressed. This is an important indicator and such metals are
thus interesting candidates for the preparation of alternative MOFs that
are less susceptible to humidity, with direct employment in
carbon-capture applications.

\section*{Acknowledgements}
This work was entirely supported by the Department of Energy Grant No.
DE-FG02-08ER46491.

\bibliography{biblio}

\end{document}